# NEW RESULT FOR THE NEUTRON $^9$Be CAPTURE AT ASTROPHYSICAL ENERGIES

*S. B. Dubovichenko*

*V. G. Fessenkov Astrophysical Institute "NCSRT" NSA RK, 050020, Almaty, Kazakhstan*
*Institute of Nuclear physics NNC RK, 050032, str. Ibragimova 1, Almaty, Kazakhstan*
dubovichenko@gmail.com

## Abstract

The total cross sections of the n$^9$Be → $^{10}$Beγ radiative capture are described in the framework of the modified potential cluster model (MPCM) with the classification of orbital states according to Young tableaux at thermal and astrophysical energies.

## 1. Introduction

The extremely successful development of microscopic models like Resonating Group Method (RGM, see, for example, [1,2]), Generator Coordinate Method (GCM, see, particularly, [3]) or algebraic version of RGM [4], leads to the view that the advancement of new results in low energy nuclear physics and nuclear astrophysics is possible only in this direction. Eventually, very common opinion states that this is the only way in which the future development of our ideas about structure of atomic nucleus, nuclear and thermonuclear reactions at low and astrophysical energies can be imagined.

However, the possibilities of simple two-cluster models were not completely studied up to now, particularly, if they use the conception of forbidden states (FS) [5] and directly take into account the resonance behavior of elastic scattering phase shifts of interacting particles at low energies [6]. But, the rather difficult RGM calculations are not the only one way for explanation of the experimental facts. It is enough to use a simple potential cluster model with the forbidden states, taking into account the classification of orbital states according to Young tableaux and the resonance behavior of the elastic scattering phase shifts. Such approach, as it was shown earlier [5-7], in many cases allows one to obtain quite adequate results in description of many experimental data.

The calculation results of the present work are interesting in terms of studies of these low energy processes, these results are of interest for the development of the primordial nucleosynthesis conception of chemical elements in the Universe, the Sun and stars [8].

First, we note that in the framework of the approach using here, notably, potential cluster model (PCM) with the classification of cluster states according to Young tableaux, with states forbidden by the Pauli principle. Earlier, there were considered 21 capture reactions of protons, neutrons and light clusters on 1$p$-shell nuclei at astrophysical energies [7,9-11] by this method. Such a model may be called a modified PCM. Further, we will consider the n$^9$Be → $^{10}$Beγ reaction at thermal and astrophysical energy range in the study of the neutron radiative capture reactions on light atomic nuclei in the MPCM with forbidden states (FS) [5,7,12]. The information about n$^9$Be interaction potentials in continuous and discrete spectra [7,13] is necessary for calculating the total cross-sections of this reaction in the frame of the MPCM [5,7]. As before, we consider that these potentials should correspond to the classification of the cluster state according to orbital symmetries [5], as was done in our works [9-11,14] on other nuclear systems taking part in different thermonuclear processes or reactions of primordial nucleosynthesis [8-7].

## 2. Classification of the orbital states

First, we draw attention to the classification of the cluster orbital states on the basis of Young tableaux for $^9$Be. If it is possible to use tableaux {44} and {1} in the 8 + 1 particle system, then in the issue of exterior product {44} × {1}, two possible orbital symmetries, {54} + {441}, will be obtained for $^9$Be. The first is forbidden as it contains five cells in one row, and the second is ranked as the allowed Young tableau, which corresponds to the allowed state of the cluster relative motion of n$^8$Be in $^9$Be [15]. Let us note that the classification of orbital states according to Young tableaux given here has only qualitative character, because for systems with $A = 9$ and 10 we cannot find tables of inner products for Young tableaux, which determine the spin-isospin symmetry of the wave function (WF) of the cluster system. These data were available for all $A < 9$ [16] and were previously used by us for the analysis of the number of allowed states (AS) and FS in wave functions for different cluster systems [7,12,17,18].

Further, the tableau {441} corresponds to the ground state (GS) of $^9$Be, so the $N^9$Be system contains FS with tableau {541} for $L = 1$ and 2 and AS with configuration {4411} with $L = 1$. Here we are limited to the minimal values of orbital moments $L$, which are required in future. The bound state in the $S$ and $D$ waves with the tableau {442} is also present. Let us note that because of the absence of tables of internal products of the Young tableaux for $A > 8$, it is impossible to unambiguously maintain – is this state forbidden or allowed? Therefore, to fix the idea we will consider that there are bound AS in the $^3S$ wave with {442}, and in the $^3D$ wave, which contains bound FS for tableau {541}, the tableau {442} is unbound. Since, the GS of $^{10}$Be with $J = 0^+$ can be formed only in the triplet spin state $^3P_0$, then further for all BS in the n$^9$Be channel we accept $S = 1$.

Thereby, N$^9$Be potentials in the $^3P_0$ wave should have forbidden and allowed bound states (BS) with Young tableaux {541} and {4411}, first of them is forbidden and the second is allowed – it corresponds to the GS of $^{10}$Be in the n$^9$Be channel. We will consider the AS for {4411} in the other $^3P$ waves as unbound, that is, their potentials will contain only one bound FS with tableau {541}. The potential, in which we match the resonance state of nucleus at the energy of 5.9599 MeV for $J^\pi = 1^-$ relatively to the GS of $^{10}$Be and bound in the n$^9$Be channel, was considered as a variant for the potential of the $^3S_1$ scattering wave with one BS for tableau {442}. Thus, we unambiguously fix the structure of FS and AS in each partial potential for $L = 0, 1, 2$, which were considered further. The similar situation was observed earlier in work [11] for n$^{12}$C and n$^{13}$C systems in the radiative neutron capture processes. Note that the number of BS, forbidden or allowed, in any partial potential determines the number of WF nodes at low distances, usually less than 1 fm [5]. It will be recalled that the WF of the BS with the minimal energy does not have nodes; the next BS has one node etc.

In our previous work [19], on the basis of consideration of the transitions only to the GS $^{10}$Be from the $^3S$ scattering wave with the zero phase shift, it was shown that the considered ambiguity of the number of forbidden or allowed BS in the $^3S_1$ scattering wave and the $^3P_0$ potential of the GS of $^{10}$Be practically does not influence on the calculation results, if these potentials contains 1-2 BS in the first case or from 2 to 3 BS in the second. In the first case, the state with the maximum energy corresponds to the tableau {442} and is allowed, in the second case, such state is allowed also and corresponded to the GS of $^{10}$Be in the n$^9$Be channel with the tableau {4411}. Therefore, the potentials of the $^3S$ and $^3P_0$ waves can be matched with the given above classification of FS and AS according to Young tableaux, considered in [19] as the second variant.

Further, we will consider transitions from the $^3D$ scattering wave, and as a ground potential we accept the variant with one FS with tableau {541}, considering that unbound AS corresponds to tableau {442}. The results of calculations with this potential are compared with two other its variants. At first place, we are considering the variant of the potentials without BS at all, which is not agree with the given above classification. As the second variant, the potential with two BS, first of them is forbidden with tableau {541}, and the second is allowed and bound with {442}. Such variant is also matched with the given above classification according Young tableaux, because it does not give an opportunity to identify – is the AS bound or not?

It is necessary keep in mind that for the $S$ wave the internal part of the nucleus is "transparent" due to absence of the Coulomb and angular momentum barrier at the neutron capture and the number of nodes in WF at low distances plays appreciable role – results for WF without nodes and with one-two nodes differ from each other noticeably [19]. The angular momentum barrier exists in the case of the $D$ wave, therefore the dependence of results from the structure of the $D$ wave in the internal range, that is, from the number of nodes at low distances, will be noticeably weaker than in the previous case.

## 3. Potential description of the scattering phase shifts

The intercluster interaction potentials of the n$^9$Be system, as usual, are chosen in simple Gaussian form [5,7]:

$$V(r) = -V_0\exp(-\alpha r^2), \qquad (1)$$

where $V_0$ and $\alpha$ are the potential parameters usually obtained on the basis of description of the elastic scattering phase shifts at certain partial waves taking into account their resonance behavior or spectrum structure of resonance levels and, in the case of discrete spectrum, on the basis of description of the BS characteristics of the n$^9$Be system. In both cases such potentials contain BS, which satisfy the classification of FS and AS according to Young tableaux given above.

Now, we would like to present more details on the construction procedure of the intercluster potentials used here, defining the criteria for finding these parameters and the order in which they are found. Primarily, the parameters of the GS potential, which are determined by the given number of allowed and forbidden states in this partial wave, are fixed quite unambiguously by the binding energy, the charge radius of the nucleus, and the asymptotic constant (AC). The accuracy of determination of the GS potential, in the first place, is connected with the AC accuracy, which is usually equal to 10–20% [7]. There are no other ambiguities in this potential, because the classification of the states according to Young tableaux allows us unambiguously fix the number of BS, which completely determines its depth, and the width of the potential fully depends on the AC value. Consequently, we obtain completely unambiguous potential with errors of parameters, determined by the spread of AC values.

The intercluster potential of the non-resonance scattering process for the case of the $^3S_1$ wave, constructed according to the scattering phase shifts at the given number of allowed and forbidden BS in the considered partial wave, is also fixed quite unambiguously. The accuracy of determination of this potential is connected, in the first place, with the accuracy of the derivation of the scattering phase shift from the experimental data and is usually equal to about 20–30%. It is difficult to estimate the accuracy with which parameters of the scattering potential are found during its construction according to the nuclear spectrum data in certain channels even at the given number of BS, although apparently it can be hoped that the error will not be much bigger than in the previous case.

We have not had any success in the data search for the n$^9$Be elastic scattering phase shifts at astrophysical energies [20], and therefore the $^3S_1$ potential of the scattering process that leads to zero scattering phase shifts at energies up to 1 MeV will be considered here. This follows from the data of the energy level spectrum of $^{10}$Be, which does not contain S-resonances with $J^\pi = 1^-$ in this energy range [21,22].

The potential is constructed completely unambiguously with the given number of BS and with the analysis of the resonance scattering when in the considered partial wave at the energies up to 1 MeV there is a rather narrow resonance with a width of about 10–50 keV. The error of its parameters does not usually exceed the error of the width determination at this level and equals 3–5%. The depth of the potential at the given number of BS completely depends on the resonance energy, and its width is determined by the width of this resonance.

Now let us go to the direct construction of the intercluster interactions – the potential of the $^3P_0^1$ ground bound state is constructed on the basis of description of the nuclear characteristics of $^{10}$Be in the n$^9$Be system, specifically the binding energy, charge radius, and AC [21]

$$V_0 = 363.351572 \text{ MeV and } \alpha = 0.4 \text{ fm}^{-2}. \qquad (2)$$

The binding energy of –6.812200 MeV, the mean square charge radius of 2.53 fm, and the mass radius of 2.54 fm were obtained with this potential. The experimental value for the charge radius of $^{10}$Be is absent in works [21-23] and for $^9$Be it equals 2.518(12) fm [23]. Later we will consider that the $^{10}$Be radius should not have a large excess over the $^9$Be radius. We are assuming that the charge neutron radius equals zero, because small deviations from zero do not play a principal role here. Its mass radius is equal to the proton radius of 0.8775(51) fm [24]. The asymptotic constant, calculated according to the Whittaker functions [25], is equal to $C_W = 1.73(1)$ at the distance 4–16 fm. The AC error is obtained by averaging it over the stated interval, where the asymptotic constant remains practically stable. Besides the allowed BS corresponding to the ground state of $^{10}$Be with {4411}, such $^3P_0$ potential has the FS with {541} in full accordance with the second variant of the classification of orbital states of clusters in the system of 10 particles in the 9+1 channel given above.

Let us cite the results of work [26] for comparison of the asymptotic constants, where $C^2 = 1.69(15)$ fm$^{-1}$ was obtained. It is necessary to note that for the definition of the asymptotic constant in that work the expression $\chi_L(R) = CW_{-\eta L+1/2}(2k_0R)$ was used, but in our calculations $\chi_L(R) = \sqrt{2k_0}\, C_0 W_{-\eta L+1/2}(2k_0R)$ [17,25]. The result of AC from work [26] is obtained for a neutron spectroscopic $S_n$ factor that is not equal to unity, as it is accepted in the present calculations. Therefore, the initial value should be divided by the value $S_n$ in this channel [26], which is equal to 0.2 according to results of theoretical work [27]. However, the value of 0.92 is given in the review [21] from the analysis of experimental data of the $^9$Be($^2$H, p$_0$)$^{10}$Be reaction for $S_n$ of $^{10}$Be in the n$^9$Be channel. Since we did not find other results for the value of $S_n$, it will be logical to accept that its average value is equal to 0.56, which in dimensionless form with $\sqrt{2k_0} = 1.045$ for AC [26] in our definition takes $C_0 = 1.66(7)$. Thus, it can be considered that the GS potential of $^{10}$Be in the n$^9$Be channel was constructed on the basis of description of the binding energy in this cluster channel and the channel AC value.

The variational method (VM) with the expansion of the cluster wave function of the relative motion of the n$^9$Be system on a non-orthogonal Gaussian basis [12,28] was used for additional control of the binding energy calculations:

$$\Phi_L(r) = \frac{\chi_L(r)}{r} = Nr^L \sum_i C_i \exp(-\beta_i r^2),$$

where $\beta_i$ and $C_i$ are the variational parameters and expansion coefficients, $N$ is the normalization coefficient of WF.

With the dimension of the basis $N = 10$, an energy of $-6.812193$ MeV was obtained for the ground state potential (2); this result differs from the finite-difference (FDM) value given above by only 7.0 eV [28]. The residuals are of the order of $10^{-10}$, the asymptotic constant in the range 5–12 fm is equal to 1.73(2), and the charge radius does not differ from the previous results [28]. The expansion parameters of the obtained variational ground state radial wave function of $^{10}$Be in the n$^9$Be cluster channel are shown in Table 1.

Table 1. The coefficients and expansion parameters of the radial variational wave function of the ground state of $^{10}$Be for the n$^9$Be channel (2) on a non-orthogonal Gaussian basis [12,28]$^{+)}$.

| $i$ | $\beta_i$, fm$^{-2}$ | $C_i$, no dim. |
|---|---|---|
| 1 | 3.243377804018342E-002 | −2.109602538815203E-003 |
| 2 | 7.917555096623996E-002 | −2.750811638982709E-002 |
| 3 | 1.784595811352628E-001 | −1.453439750665609E-001 |
| 4 | 3.746612146329068E-001 | −4.789696250511998E-001 |
| 5 | 1.126366082891110 | 2.312966715268185 |
| 6 | 2.331534330644200 | 3.209514146107344 |

| 7 | 2.741502097647982 | −5.201076539985449 |
| 8 | 3.216691148332270 | 3.694013806676463 |
| 9 | 3.852002896536004 | −1.161048304949317 |
| 10 | 4.956980070114872 | 1.186950441156130E-001 |

+) The normalization coefficient of the wave function in the interval of 0–25 fm is $N = 1.000000000000000$ fm$^{-L}$.

Since the variational energy decreases as the dimension of the basis increases and yields the upper limit of the true binding energy, and the finite-difference energy increases as the step size decreases and number of steps increases [23], the average value of −6.8121965(35) MeV can be taken as a realistic estimate of the binding energy in this potential. Therefore, the real accuracy of determination of the binding energy of $^{10}$Be in the n$^9$Be cluster channel for this potential, using two different methods (FDM and VM) and two different computer programs for the potential (2), is at the level of ±3.5 eV.

For the potential of the first $^3P_2^1$ excited state (ES) of $^{10}$Be at the energy of −3.44417 MeV in the n$^9$Be channel with $J^\pi = 2^+$ the parameters are obtained as:

$$V_0 = 345.676472 \text{ MeV and } \alpha = 0.4 \text{ fm}^{-2}. \quad (3)$$

This potential results in an energy of −3.444170 MeV, a charge radius of 2.54 fm, a mass radius of 2.57 fm, and an AC equal to 1.15(1) in the interval 4–18 fm, and has one bound FS at {541}; that is, it applies to the given classification of the orbital cluster states according to Young tableaux. The state with {4411} is considered here as unbound. Since we have no data on AC of the ES, then the width of these potentials $\alpha$ is used for these potentials, which for the GS potential provides the true value of its AC.

The next parameters,

$$V_0 = 328.584413 \text{ MeV and } \alpha = 0.4 \text{ fm}^{-2}, \quad (4)$$

were obtained for the potential of the second excited state of $^{10}$Be at the energy of −0.85381 MeV in the n$^9$Be channel with $J^\pi = 2^+$, which also has one FS.

This potential allows us to obtain a binding energy of −0.853810 MeV, a charge radius of 2.55 fm, a mass radius of 2.69 fm, and an AC equal to 0.60(1) in the interval 4–26 fm.

Further, let us note that the fourth $^3P_0^2$ excited level at the energy of 6.17930(7) MeV relative to the GS or −0.6329 MeV relative to the threshold of the n$^9$Be channel of $^{10}$Be [21] coincides with the GS with respect to its quantum numbers $J^\pi T = 0^+1$. Therefore, it is possible to consider the $E1$ transition to this BS from the $S$ scattering wave. The potential of this BS has the parameters

$$V_0 = 326.802239 \text{ MeV and } \alpha = 0.4 \text{ fm}^{-2}, \quad (5)$$

and leads to a binding energy in this channel of −0.632900 MeV, a charge radius of 2.56 fm, a mass radius of 2.72 fm, and an AC equal to 0.53(1) in the interval 4–28 fm.

Besides, the third $^3S_1$ excited level with $J^\pi T = 1^-1$ [21] is observed at the energy of 5.95990(6) MeV or –0.8523 MeV relative to the threshold of the n$^9$Be channel. Therefore, it is possible to consider the $M1$ transition from the $^3S_1$ non-resonance scattering wave to the same $^3S_1$ bound state of $^{10}$Be. However, it is usually considered that the cross-sections of such a $^9$Be(n,γ$_3$)$^{10}$Be process will be lower by one to two orders of magnitude than those obtained in the $E1$ capture.

Pay attention again that, similarly to the n$^{12}$C system [11], the third ES of $^{10}$Be with $J^\pi = 1^-$ at 5.9599 MeV could be considered as bound AS in the $^3S_1$ scattering wave with tableau {422}. Therefore, we use for the $^3S_1$ wave the potential of the form:

$$V_0 = 33.768511 \text{ MeV and } \alpha = 0.4 \text{ fm}^{-2}, \qquad (6)$$

which leads to a binding energy of –0.852300 МэВ [21], charge and mass radii of 2.57 fm and 2.76 fm, and an AC equal to 1.24(1) in the interval 4–30 fm. The phase shift of this potential is shown in Fig. 1 by the dashed line and smoothly decreases till 121° at 1.0 MeV.

Such potential does not contain FS and will be used for calculations of the total cross sections of the $E1$ transition from the $^3S_1$ scattering wave to the $^3P_0^1$ GS of $^{10}$Be in the n$^9$Be channel and fourth $^3P_0^2$ excited state at 6.1793 MeV with $J^\pi = 0^+$, that is, $^3S_1 \to {}^3P_0^1 + {}^3P_0^2$. In addition, the transition to the first ES at the energy of 3.36803 MeV and to the second $^3P_2^2$ excited state at 5.95839 MeV with $J^\pi = 2^+$, that is, $^3S_1 \to {}^3P_2^1 + {}^3P_2^2$.

Further, let us consider the resonance states of the n$^9$Be scattering at the energy lower than 1 MeV. It is known that in the spectrum of $^{10}$Be in the n$^9$Be channel there is superthreshold level with $J^\pi = 3^-$ at the energy of 0.6220(8) MeV in laboratory system (l.s.) and width of 15.7 keV [21], which can be matched to the $^3D_3$ resonance in the n$^9$Be elastic scattering. The level with $J^\pi = 2^+$ at the energy of 0.8118(7) MeV (l.s.) and width of 6.3 keV, which can correspond to the $^3P_2$ resonance in the n$^9$Be scattering [21,22], is also observed. In the first case, the $E3$ transition has to be considered and in the second the $M2$ transition to the GS of $^{10}$Be, whose cross-sections are appreciably lower than those of the $E1$ process and will not be taken into account in future.

However, it is possible to consider the $E1$ transition, for example, from the $^3D_3$ resonance scattering wave to the first and second excited states of $^{10}$Be with $J^\pi = 2^+$ at the energies of 3.36803 MeV and 5.95839 MeV relative to the GS, which are bound at –3.44417 MeV and –0.85381 MeV relative to the threshold of the n$^9$Be channel [21]. All calculated cross-sections of the $^9$Be(n,γ)$^{10}$Be reaction will be smooth due to this resonance only in the energy range up to 0.1-0.2 MeV.

The next parameters were found for the $^3D_3$ scattering wave

$$V_0 = 457.877 \text{ MeV and } \alpha = 0.35 \text{ fm}^{-2}. \qquad (7)$$

The phase shift of this potential is shown in Fig. 1 by the solid line and has a resonance character, and the potential itself contains the bound FS with {541} tableau in accordance with the classification given above, and the state with {442} is not bound. This variant of the potential, with this structure of FS, will be considered as the main, but, for comparison, we will consider further another two variants.

If one uses the expression [29] for the calculation of the level width by the phase shift δ,

$$\Gamma = 2(d\delta/dE)^{-1},$$

then the width of this resonance is equal to 15.0 keV (c.m. system), which is in quite good agreement with the results of work [21]. The second variant of the potential has the parameters

$$V_0 = 132.903 \text{ MeV and } \alpha = 0.22 \text{ fm}^{-2}. \tag{8}$$

The phase shift of this potential is shown by the dot-dashed line in Fig. 1, the resonance is also at 622 keV with a width of 15.7 keV, and the potential itself does not contain bound forbidden or allowed states; that is, this potential corresponds to the case when FS for {541} is absent, and the state with {442} is not bound.

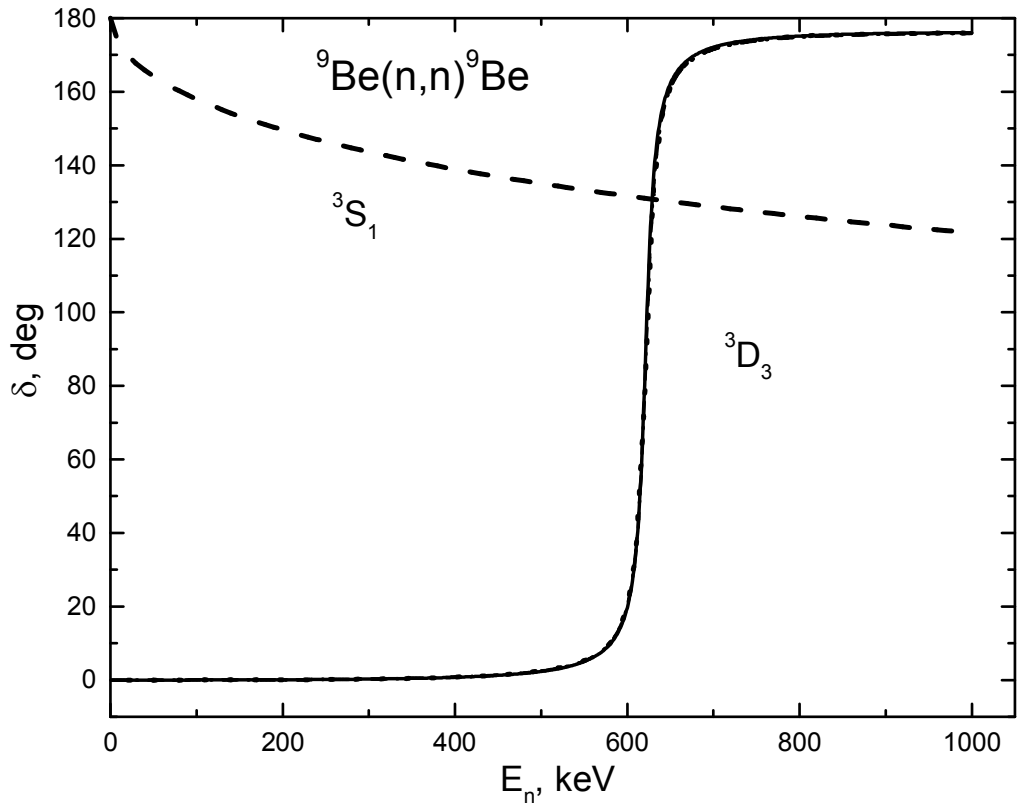

Fig. 1. The $^3S_1$ and $^3D_3$ phase shifts of the n$^9$Be elastic scattering at low energies. The lines show the calculations with the Gaussian potential, whose parameters are given in the text.

Now, we will consider the variant of the $^3D_3$ scattering potential with the parameters:

$$V_0 = 985.183 \text{ MeV and } \alpha = 0.43 \text{ fm}^{-2}, \qquad (9)$$

which leads to the width of 15.5 keV at the resonance of 622 keV and has two bound states. First of them corresponds to the bound FS with {541}, second – to the bound state at {442}. The scattering phase shift is shown in Fig. 1 by the dotted line, which differs from the solid line only at energies lower than 400 keV.

The next parameter values are used for the potentials of the $^3D_2$ and $^3D_1$ waves:

$$V_0 = 300.0 \text{ MeV and } \alpha = 0.35 \text{ fm}^{-2}, \qquad (10)$$

which lead to the zero phase shifts, because there are no resonances in these waves. This potential contains the bound FS with {541} in accordance with the given above classification, and the state {442} is not bound.

As it was said above, the resonance level with $J^\pi = 2^+$ at the energy of 0.8118(7) MeV (l.s.) above the threshold of the n$^9$Be channel and with the width of 6.3 keV that can be matched to the $^3P_2$ resonance [21,22], is considered in the n$^9$Be elastic scattering. There are no resonances in the other partial $^3P_0$ and $^3P_1$ waves, and their potentials containing one bound FS for tableau {541} have to lead to near-zero phase shifts

$$V_0 = 206.0 \text{ MeV and } \alpha = 0.4 \text{ fm}^{-2}. \qquad (11)$$

We have not had any success in finding the potential for the resonance $^3P_2$ wave, which is able to correctly describe its small width of 6.3 keV [21]. Therefore, further, the identical and given above potential (11), was used for calculations of the transitions $^3P_0 + ^3P_1 + ^3P_2 \to ^3S_1$ for all $^3P$ waves, that is, we are not taking into account the influence of the $^3P_2$ resonance.

Accurate values of particle masses are specified in our calculations [24,30]; the $\hbar^2/m_0$ constant, where $m_0$ is the atomic unit, is equal, as usual, to 41.4686 MeV fm$^2$. Although it is currently considered that this value is slightly out of date, we are continuing to use it for ease of comparison of the new results with all previous results (see, for example, [5,7] and [17,31]).

## 4. Total capture cross-sections

The expressions for the total radiative capture cross-sections $\sigma(NJ,J_f)$ in the potential cluster model are given, for example, in works [17] or [32,33] and are written as

$$\sigma_c(NJ,J_f) = \frac{8\pi K e^2}{\hbar^2 q^3} \frac{\mu}{(2S_1+1)(2S_2+1)} \frac{J+1}{J[(2J+1)!!]^2} \times A_J^2(NJ,K) \sum_{L_i,J_i} P_J^2(NJ,J_f,J_i) I_J^2(J_f,J_i)$$

where for the electric orbital $EJ(L)$ transitions ($S_i = S_f = S$) we have [32]:

$$P_J^2(EJ,J_f,J_i) = \delta_{S_i S_f}\left[(2J+1)(2L_i+1)(2J_i+1)(2J_f+1)\right](L_i 0 J 0 | L_f 0)^2 \begin{Bmatrix} L_i & S & J_i \\ J_f & J & L_f \end{Bmatrix}^2,$$

$$A_J(EJ,K) = K^J \mu^J \left(\frac{Z_1}{m^J_1} + (-1)^J \frac{Z_2}{m^J_2}\right), \qquad I_J(J_f, J_i) = \langle \chi_f | R^J | \chi_i \rangle.$$

Here, $q$ is the wave number of the initial channel particles; $L_f$, $L_i$, $J_f$, $J_i$, $S_f$, and $S_i$ are the angular moments of particles in the initial (*i*) and final (*f*) channels; $S_1$ and $S_2$ are spins of the initial channel particles; $m_1$, $m_2$, $Z_1$, and $Z_2$ are the masses and charges of the particles in the initial channel, respectively; $K$ and $J$ are the wave number and angular moment of the γ-quantum in the final channel; and $I_J$ is the integral over wave functions of the initial $\chi_i$ and final $\chi_f$ states as functions of relative cluster motion with the intercluster distance $R$.

The $E1$ process from the $^3S_1$ non-resonance scattering wave with the potential (6) to the $^3P^1_0$ bound ground state of $^{10}$Be with $J^\pi T = 0^+1$ for the potential (2), that is, $^3S_1 \to {^3P^1_0}$ transition, was taken into account in the consideration of the electromagnetic transitions in the n$^9$Be cluster channel. The calculation results of the total capture cross-sections are shown in Fig. 2 by the dot-dot-dashed line for these potentials. The experimental data for the total cross-sections at energies from 25 meV to 25 keV (l.s.) were taken from works [34-36], respectively.

The dashed line in Fig. 2 gives the results for the total cross-sections of the transition $^3S_1 \to {^3P^1_2}$ with the combination of the potentials of the first ES in the $^3P_2$ wave (3) and the $^3S_1$ scattering wave (6). The dot-dashed line shows the results for the transition $^3S_1 \to {^3P^2_2}$ with potentials of the second ES (4) and the $^3S_1$ wave (6). The dotted line shows the results for the transition from the $^3S$ scattering wave (6) to the fourth $^3P^2_0$ ES with $J=0^+$ with the potential (5), and the solid line is the sum of all these transitions.

Now we will consider the $E1$ transitions from the $^3D_3$ resonance scattering wave with $J^\pi T = 3^-1$ at the energy of 0.6220(8) MeV (l.s.) and width of 15.7 keV [21] to the first and second excited states of $^{10}$Be with $J^\pi T = 2^+1$, which can be compared with $^3P_2$ bound states of the n$^9$Be system.

The calculation results of the total cross-section of the $E1$ transition to the second excited state with potentials (4) and (7) are shown in Fig. 3 by the dotted line, while the dashed line denotes the transition to the first excited state with potentials (3) and (7). The dot-dashed line gives the sum of all cross-sections of the $E1$ transition, shown in Fig. 2 by the solid line. The solid line in Fig. 3 is the sum of all cross-sections of the considered transitions in the $^9$Be(n, γ)$^{10}$Be reaction. The transition total cross-section at the resonance energy of 622 keV from the $^3D_3$ scattering wave to the first excited state at the energy of –3.44417 MeV reaches 2.45 mb and that to the second one at the energy of –0.853810 MeV equals approximately 0.16 mb.

Also, we have considered the transitions from the $^3D_2$ or $^3D_1$ scattering states with potentials (10) to the first ES for the potential (3) $^3D_1 + {^3D_2} \to {^3P_2}$ and the transition from the $^3D_1$ scattering state to the GS of $^{10}$Be with the potential (2), that is, $^3D_1 \to {^3P^1_0}$. In the first case, the capture cross section is shown by close dots in the right part of Fig. 2. There is practically no difference from the results of second case with the transition to the GS. Hence it follows that such transitions become play a visible role only at energies up to 1 MeV.

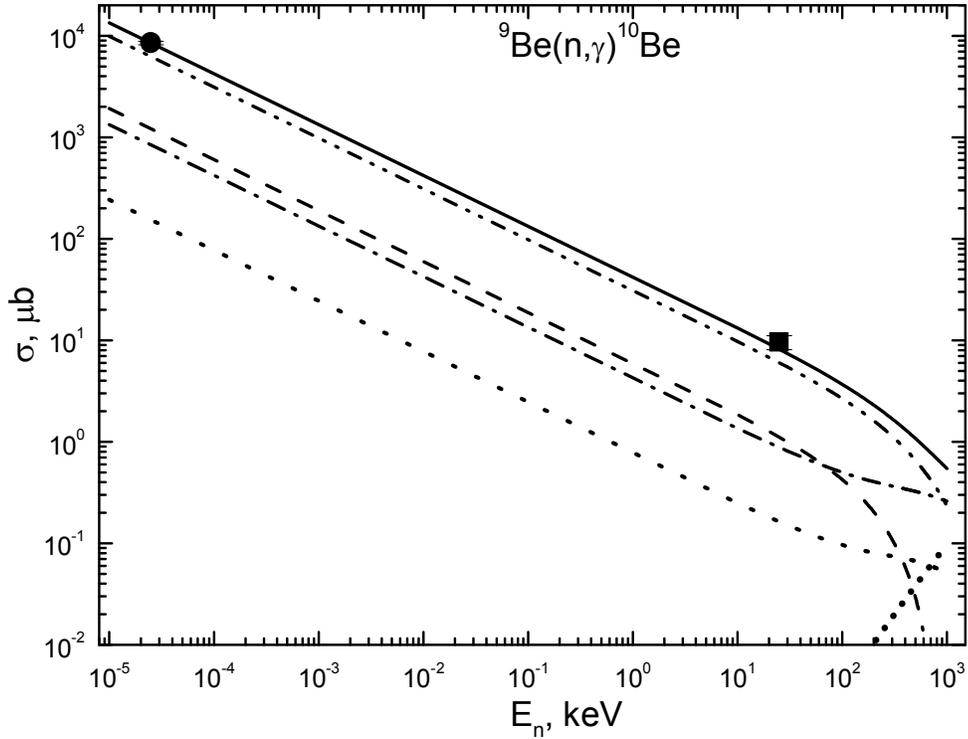

Fig. 2. The total cross-sections of the radiative neutron capture on $^9$Be. Points (●) indicate experimental data from work [34] at 25 meV and squares (■) indicate experimental data from [35,36] at 25 keV. Lines show the total cross-sections calculation results.

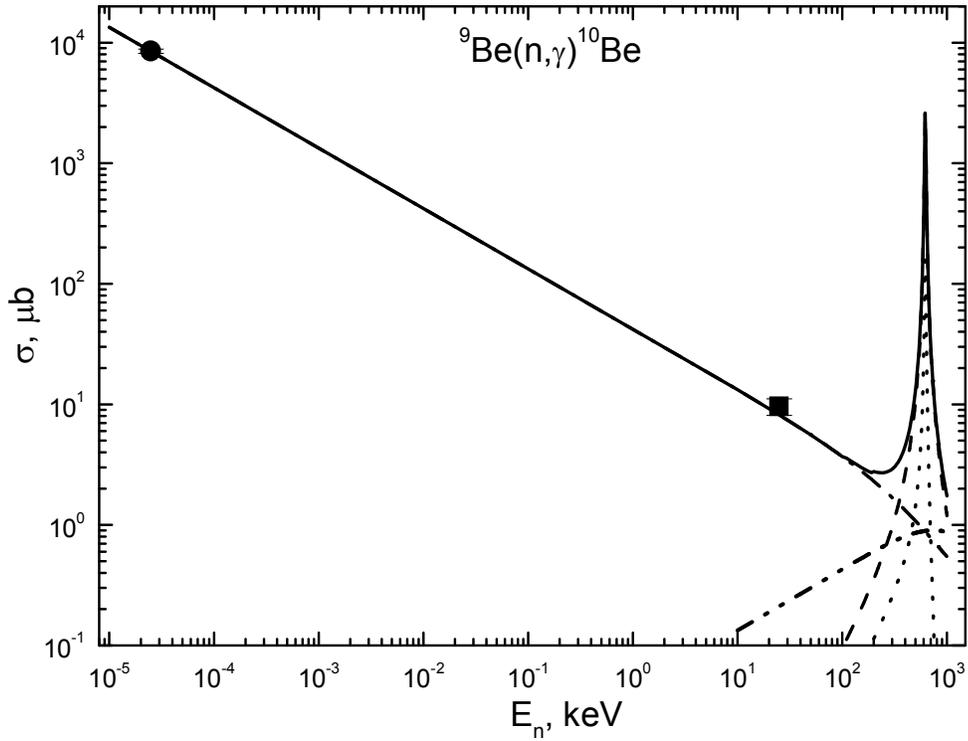

Fig. 3. The total cross-sections of the radiative neutron capture on $^9$Be taking into account the resonance at 622 keV. Points (●) represent experimental data from work [34] at 25 meV and squares (■) represent experimental data from [35,36] at 25 keV. Lines show the total cross-sections calculation results.

The results of calculations of $^9Be(n, \gamma_3)^{10}Be$ for the sum of all three $E1$ transitions $^3P_0 + ^3P_1 + ^3P_2 \rightarrow ^3S_1$ with potentials of all $P$ waves (11) and BS (6) are given in Fig. 3 by the dot-dot-dashed line. It is seen that the cross-sections, obtained in the issue of such transitions, do not give essential contribution to the summarized cross-sections only at the energy up to 10 keV, and from the range above 150 keV their value exceeds cross-sections for the $^3S_1 \rightarrow ^3P_0^1$ transition, shown in Fig. 3 by the dot-dashed line.

Since at the energies from $10^{-5}$ keV up to 10 keV the calculated cross-section shown in Fig. 2 by the solid line is almost a straight line, it can be approximated by a simple function of the form:

$$\sigma_{ap}(\mu b) = \frac{42.4335}{\sqrt{E_n(keV)}}.$$

The value of constant 42.4335 $\mu b \cdot keV^{1/2}$ was determined from a single point of the cross-sections with a minimal energy of $10^{-5}$ keV. Further, it is possible to consider the absolute value of the relative deviation of the calculated theoretical cross-sections and the approximation of this cross-section by the expression given above in the energy range from $10^{-5}$ to 10 keV:

$$M(E) = \left|[\sigma_{ap}(E) - \sigma_{theor}(E)] / \sigma_{theor}(E)\right|.$$

It was found that this deviation does not exceed 0.7% at energies lower than 10 keV, making it possible to use the cross-section approximation given above in the majority of applied problems. It appears to be possible to suppose that this form of total cross-section dependence on energy will have conserved at lower energies. In this case, estimation of the cross-section value, for example at the energy of 1 $\mu$ keV, gives the value of 1.3 b.

## 5. Conclusion

Thus, it is possible to acceptably describe the available experimental data on the total cross-sections of neutron capture at the energies at 25 meV and 25 keV if certain assumptions are made concerning the general character relative to the interaction potentials in the $n^9Be$ channel of $^{10}Be$.

In the case of consideration of the $E1$ transition from the $^3D_3$ resonance scattering wave at the energy of 622 keV to the first and second excited states of $^{10}Be$ in the capture cross-sections, the observed resonance reaches about 2.6 mb. The usage of the potentials for the $^3D_3$ resonance phase shift with different numbers of FS has almost no effect on the resonance value, which influences the cross-sections at energies above 0.2 MeV. Therefore, it always can be considered that the $^3D$ waves contain one bound FS with tableau {541} and the state with {442} is unbound. At the same time the $^3S_1$ wave contains one allowed BS with tableau {442}, and FS are absent. All $^3P$ potentials contain one bound FS with tableau {541}, and $^3P_0$ wave includes another allowed BS, corresponding to the GS of $^{10}Be$. Thus, all considered potentials quite correspond to the given above classification of FS and AS according to Young tableaux, and allow one to correctly describe available experimental data under consideration of few possible $E$1 transitions between different levels of $^{10}$Be.

We should note that the experimental data on the cross-sections of the considered reaction that are available for our disposal are obviously insufficient. Further, more thorough consideration of the $n^9\text{Be} \to {}^{10}\text{Be}\gamma$ radiative capture at thermal and astrophysical energies is apparently needed, especially in the range of the resonance with $J^\pi = 3^-$ at 0.622 MeV.

**Acknowledgments**

This work was supported by the Grant Program No. 0151/GF2 of the Ministry of Education and Science of the Republic of Kazakhstan "The study of thermonuclear processes in the primordial nucleosynthesis of the Universe".

Author would like to express thanks to Professor R. Yarmukhamedov for the detailed discussions of some questions of the work and for the provision of his results on the asymptotic normalization constants.

Author also would like to express thanks to A.V. Dzhazairov-Kakhramaniov for their great help in translating this paper.

**References**

1. Wildermut, K., & Tang, Y. C. 1977, A unified theory of the nucleus (Branschweig: Vieweg)
2. Mertelmeir, T., & Hofmann, H. M. 1986, Nucl. Phys. A, 459, 387
3. Descouvemont, P., & Dufour, M. 2012, Microscopic cluster model, in: Clusters in Nuclei, Vol. 2, Lecture Notes in Physics 848, ed. by C. Beck. (Berlin: Springer-Verlag Berlin Heidelberg), 1
4. Nesterov, A. V., et al., 2010, Phys. Part. Nucl., 41, 716
5. Nemets, O. F., Neudatchin, V. G., Rudchik, A. T., Smirnov, Y. F., Tchuvil'sky, Yu. M. 1988, Nucleon Association in Atomic Nuclei and the Nuclear Reactions of the Many Nucleon Transfers (Kiev: Naukova dumka) (in Russian)
6. Dubovichenko, S. B. 2011, Thermonuclear processes in the Universe, Series: Kazakhstan space research, Vol. 7 (2nd ed.; Almaty: A-tri), arXiv:1012.0877 [nucl-th]; Dubovichenko, S. B. 2012, Thermonuclear Processes of the Universe (New-York: NOVA Scientific Publishing), https://www.novapublishers.com/catalog/product_info.php?products_id=31125
7. Dubovichenko, S. B. 2012, Selected methods of nuclear astrophysics, (2nd ed.; Saarbrucken, Germany: Lambert Acad. Publ. GmbH&Co. KG), https://www.lap-publishing.com/catalog/details/store/gb/book/978-3-8465-8905-2/Избранные-методы-ядерной-астрофизики
8. Barnes, C. A., Clayton, D. D., & Schramm, D. N. 1982, Essays in Nuclear Astrophysics (Cambridge, UK: Cambridge University Press)
9. Dubovichenko, S. B. & Dzhazairov-Kakhramanov, A. V. 2012, Ann. der. Phys., 524, 850
10. Dubovichenko, S. B., & Dzhazairov-Kakhramanov, A. V. 2012, Int. J. Mod. Phys. E, 21, 1250039-1

11. Dubovichenko, S. B., Dzhazairov-Kakhramanov, A. V. & Burkova, N. A. http://arxiv.org/abs/1202.1420

12. Dubovichenko, S. B. 2004, Properties of the light nuclei in potential cluster model (Almaty: Daneker), http://arxiv.org/abs/1006.4944

13. Dubovichenko, S. B., Takibaev, N. Zh., & Chechin, L. M. 2008, Physical Processes in the Far and Near Space, Series "Kazakhstan space research", Vol. 3 (Almaty: Daik-Press), http://www.arxiv.org/abs/1012.1705

14. Dubovichenko, S. B. 2012, Rus. Phys. J. 55, 138

15. Neudatchin, V. G. & Smirnov, Yu. F. 1969, Nucleon associations in light nuclei (Moscow: Nauka)

16. Itzykson, C., & Nauenberg, M. 1966, Rev. Mod. Phys., 38, 95

17. Dubovichenko, S. B. & Dzhazairov-Kakhramanov, A. V. 1997, Phys. Part. Nucl., 28, 615

18. Dubovichenko, S. B. & Uzikov, Yu. N. 2011, Phys. Part. Nucl., 42, 251

19. Dubovichenko, S. B., & Burkova, N. A. 2013, Rus. Phys. J., 56 (In press)

20. http://cdfe.sinp.msu.ru

21. Tilley, D. R., et al., 2004, Nucl. Phys. A, 745, 155

22. Ajzenberg-Selove, F. 1988, Nucl. Phys. A, 490, 1

23. http://cdfe.sinp.msu.ru/cgi-bin/muh/radchartnucl.cgi?zmin=0&zmax=14&tdata=123456

24. http://physics.nist.gov/cgi-bin/cuu/Category?view=html&Atomic+and+nuclear.x=78&Atomic+and+nuclear.y=12

25. Plattner, G. R., & Viollier, R. D. 1981, Nucl. Phys. A, 365, 8

26. Dolinskii, E. I., Mukhamedzhanov, A. M., Yarmukhamedov, R. 1978, Direct nuclear reactions on light nuclei with the emission of neutrons (Tashkent: FAN)

27. Boyarkina, F. N. 1973, The structure of 1$p$-shell nuclei (Moscow: Moscow State University)

28. Dubovichenko, S. B. 2012, Calculation methods for nuclear characisttics (2nd ed.; Saarbrucken, Germany: Lambert Academy Publ.), https://www.lap-publishing.com/catalog/details//store/ru/book/978-3-659-21137-9/методы-расчета-ядерных-характеристик

29. Mott, N., & Messey, H. 1949, The Theory of Atomic Collisions (London: Oxford)

30. http://cdfe.sinp.msu.ru/services/ground/NuclChart_release.html

31. Dubovichenko, S. B., & Dzhazairov-Kakhramanov, A.V. 1990, Sov. J. Nucl. Phys. USSR, 51, 971

32. Angulo, C., et al., 1999, Nucl. Phys. A, 656, 3

33. Fowler, W., et al., 1975, Ann. Rev. Astr. Astrophys, 13, 69

34. Conneely, C. M., Prestwich, W. V., & Kennett, T. J. 1986, J. Nucl. Inst. Method. A, 248, 416

35. Wallner, A., et al., 2008, J. Phys. G, 35, 014018

36. Wallner, A., et al., 2010, Nucl. Instr. Methods B, 268, 1277